\documentclass[preprint,aps,prd,groupedaddress,showpacs,nofootinbib]{revtex4-1}
\usepackage{hyperref}
\usepackage{pifont}
\usepackage{amsmath,amsfonts,amssymb,amsthm,mathrsfs}
\usepackage{graphicx,chngcntr,vmargin}
\setmarginsrb{20mm}{10mm}{20mm}{20mm}{10mm}{10mm}{10mm}{10mm}
\newcommand{\oo}{\infty}
\renewcommand{\d}{\mathrm{d}}
\newcommand{\bea}{\begin{eqnarray}}
\newcommand{\eea}{\end{eqnarray}}

\newcommand{\Tauc}{\mathcal{T}^{\text{cut}}}
\newcommand{\Tau}{\mathcal{T}}

\makeatletter
\newcommand\footnoteref[1]{\protected@xdef\@thefnmark{\ref{#1}}\@footnotemark}
\makeatother

\begin{document}

\title{Next-to-leading-logarithmic power corrections for
  $N$-jettiness subtraction in color-singlet production}
\author{Radja Boughezal}
\affiliation{High Energy Physics Division, Argonne National Laboratory, Argonne, IL 60439, USA}
\author{Andrea Isgr\`o}
\affiliation{Department of Physics \& Astronomy, Northwestern University, Evanston, IL 60208, USA} 
\author{Frank Petriello}
\affiliation{High Energy Physics Division, Argonne National Laboratory, Argonne, IL 60439, USA}
\affiliation{Department of Physics \& Astronomy, Northwestern University, Evanston, IL 60208, USA}

\begin{abstract}

We present a detailed derivation of the power corrections to the
factorization theorem for the 0-jettiness event
shape variable $\Tau$.  Our calculation is performed directly in QCD without
using the formalism of effective field theory.  We analytically calculate the
next-to-leading logarithmic power corrections to the inclusive cross section for small $\Tau$ at next-to-leading order in
the strong coupling constant, extending previous computations which
obtained only the leading-logarithmic power corrections.  We present a numerical study of the power corrections in the context of their application to
the $N$-jettiness subtraction method for higher-order calculations,
using gluon-fusion Higgs production as an example.
The inclusion of the next-to-leading-logarithmic power corrections
further improves the numerical efficiency of the approach 
beyond the improvement obtained from the leading-logarithmic power
corrections.

\end{abstract}

\date{\today}
\maketitle
\tableofcontents

\section{Introduction}

The increasing precision of data from the Large Hadron Collider (LHC)
and the anticipated precision of potential future experiments
require calculations to increasingly high orders in
the perturbative expansion of QCD.  The calculation of higher-order
corrections is complicated by the fact that the real-emission and
virtual contributions to the cross section exhibit
infrared singularities that cancel only after they are combined.
Currently, the fully differential predictions needed to model the
final-state cuts imposed in all experimental analyses can be obtained
at next-to-next-to-leading order (NNLO) in the strong coupling
constant for $2 \to 2$ scattering
processes.  These have become available only recently due to rapid
progress in developing new schemes for the efficient cancellation of
infrared singularities~\cite{GehrmannDeRidder:2005cm, Somogyi:2005xz,
  Catani:2007vq, Czakon:2010td,
  Boughezal:2011jf,Cacciari:2015jma,Boughezal:2015dva,Gaunt:2015pea}.
With these advances a new standard in the comparison of theoretical
predictions with data from the LHC has been achieved.

We discuss here one of these successful new approaches to
higher-order QCD calculations, the $N$-jettiness subtraction
scheme~\cite{Boughezal:2015dva, Gaunt:2015pea}.  This method uses the
$N$-jettiness event shape variable $\Tau_N$~\cite{Stewart:2010tn} as a
resolution parameter to isolate and cancel the double-unresolved
singular limits, where two partons become soft and/or collinear, that
complicate the calculation of NNLO cross sections.  It has led to some
of the first calculations for vector boson production in association
with a jet~\cite{Boughezal:2015dva, Boughezal:2015ded,
  Boughezal:2016yfp,Boughezal:2016dtm,Boughezal:2016isb} and Higgs
production in association with a jet~\cite{Boughezal:2015aha} at the
LHC through NNLO.  It has also led to first predictions for inclusive
jet production at NNLO in electron-nucleon
collisions~\cite{Abelof:2016pby}.  The $N$-jettiness subtraction
scheme relies upon the introduction of a cutoff $\Tau_N^{\text{cut}}$ that
separates the $N+1$ jet region of an $N$-jet production process from the doubly-unresolved
limit.  The below-cut region is expanded in $\Tau_N^{\text{cut}}/Q$, where
$Q$ denotes the hard momentum transfer in the process, in order to
allow for an effective field theory calculation using soft-collinear
effective
theory (SCET)~\cite{Bauer:2000ew,Bauer:2000yr,Bauer:2001ct,Bauer:2001yt,Bauer:2002nz}.
A small cutoff is needed  so that the power corrections in
$\Tau_N^{\text{cut}}/Q$ are negligible.  However, the below-cut and above-cut
contributions separately depend on logarithms of $\Tau_N^{\text{cut}}/Q$ that
only cancel after combining the two regions. For small $\Tau_N^{\text{cut}}$ these logarithms
introduce numerical noise that must be controlled.  Although the
numerics can already be controlled sufficiently for phenomenological
applications, it is desirable for computational efficiency to reduce
the sensitivity of the method to the power corrections.  One approach
is to analytically calculate the power corrections
to the SCET factorization theorem used in the below-cut region.  The
study of the structure of sub-leading power corrections to
effective-theory factorization theorems is also of general interest,
and has received significant attention recently~\cite{Bonocore:2014wua,Bonocore:2015esa,Bonocore:2016awd,Moult:2017rpl,Feige:2017zci,Chang:2017atu,Beneke:2017ztn}.

The importance of and interest in the $N$-jettiness subtraction scheme
has motivated several calculations of the power corrections to the
factorization theorem for the simplest case of color-singlet
production, which uses the 0-jettiness event shape variable.  The leading-logarithmic power corrections
to the below-cut Drell-Yan cross section, which scale as $\alpha_s
\Tau_N^{\text{cut}} \log\,\Tau_N^{\text{cut}}$ and $\alpha_s^2
\Tau_N^{\text{cut}} \log^3\,\Tau_N^{\text{cut}}$ at NLO and NNLO respectively, were derived
using an effective field theory approach in
Ref.~\cite{Moult:2016fqy}. 
The leading-logarithmic power corrections
for both Drell-Yan and inclusive gluon-fusion Higgs production were derived
using both an effective-theory method and a direct QCD calculation in
Ref.~\cite{Boughezal:2016zws}.   The corrections to the differential cross section for gluon-fusion Higgs
production were subsequently considered using an effective-theory
approach in Ref.~\cite{Moult:2017jsg,Ebert:2018lzn}. 

Motivated by the increasing interest in
understanding power corrections in effective field theory, we revisit and extend the calculation of the power corrections
to the 0-jettiness factorization theorem.  We attempt to fulfill
several goals in this work.
\begin{itemize}

\item We present a detailed derivation of the leading-power cross section
  and the sub-leading power corrections. For the leading-power terms we show explicitly
  how the calculation using the SCET factorization theorem maps onto a
  direct derivation using QCD.   We study gluon-fusion Higgs production as a
  representative example.  Our aim is to provide a detailed, pedagogical discussion to
  allow interested readers with differing expertise to follow the derivations in
  detail.

\item  We present in detail an
  explicit calculation of the NLO power corrections for the inclusive cross section for gluon-fusion
  Higgs production directly in QCD, assuming an arbitrary hardness measure used to
  define $\Tau_0$.

\item In addition to rederiving in detail the leading-logarithmic
  sub-leading power corrections, we also derive analytically for the
  first time the  next-to-leading-logarithmic (NLL) $\mathcal{O} \left( \Tau_0\right)$ power corrections.

\item We study in detail the numerical impact of the LL and NLL power
 corrections for two different choices of hardness measure in the $\Tau$ definition on the NLO gluon-fusion Higgs inclusive cross section
 computed using $N$-jettiness subtraction.  The inclusion of the full
 power corrections further improves the numerical efficiency of the
 method beyond the leading-logarithmic improvements observed in
 previous work~\cite{Moult:2016fqy,Boughezal:2016zws,Moult:2017jsg}.
 Deviations from the NLO correction obtained using dipole subtraction
 remain below 1\%, and they can be greatly reduced by changing the normalization factor in the $\Tau$ definition.

\end{itemize}

Our manuscript is organized as follows.  We present our notation and
define the versions of 0-jettiness we consider in
Section~\ref{sec:taudef}.  For pedagocial purposes, and as a check on
our direct derivation of the $\Tau$ distribution including power
corrections, we derive the leading-power result using the SCET
factorization theorem in Section~\ref{sec:SCETresult}.  Our direct QCD
derivation of the leading-power result, as well as the full ${\cal
  O}(\Tau)$ power corrections, is shown in detail in
Section~\ref{sec:qcd}.  We present numerical results for two different choices of hardness measure in the $\Tau$ definition in Section~\ref{sec:numerics}.  Finally, we
conclude in Section~\ref{sec:conc}.  In an Appendix we list the
formulae for the NLL power corrections for the $qg$ and $q\bar{q}$
contributions to Higgs production.  The expressions for the dominant
$gg$ channel are given in the main text.

\section{Notation and $\Tau$ definitions}
\label{sec:taudef}

In this section we present the notation and conventions used in this
paper, and define the $N$-jettiness event-shape variable. To simplify
comparison with other results in the literature we adopt whenever
possible the notation of Ref.~\cite{Moult:2017jsg}.

The most general definition of the $N$-jettiness event-shape variable is
\begin{equation}
\Tau_N = \sum_k \min_i \left\{ \frac{2 q_i \cdot p_k}{Q_i}\right\},
\end{equation}
where the index $i$ runs over the two beam directions and over the $N$
final-state jets.  The index $k$ runs over all final-state partons.  The $Q_i$ are arbitrary hardness measures that lead
to different definitions of $N$-jettiness.  For our gluon-fusion Higgs
production example we will be interested in
0-jettiness in the situation where there is only a single parton in the
final state.  In this case the relevent event shape variable is
\begin{equation}
\Tau_0 = \min \left\{ \frac{2 q_a \cdot p_3}{Q_a}, \frac{2 q_b \cdot p_3}{Q_b} \right\},
\label{eq:bigtau}
\end{equation}
where $p_3$ is the momentum of the emitted final-state parton.
The two beam momenta $q_a$ and $q_b$ are
\begin{equation}
\label{eq:qdef}
q_a= \frac{x_a \sqrt{s}}{2} \, \, n^\mu, \qquad \qquad q_b  = \frac{x_b \sqrt{s}}{2} \, \, \overline{n}^\mu,
\end{equation}
where $x_a$ and $x_b$ are the momentum fraction carried by the two partons, and
\begin{equation}
n^\mu \equiv \begin{pmatrix} 1\\0\\0\\1\end{pmatrix}, \qquad \qquad
\overline{n}^\mu \equiv \begin{pmatrix} 1\\0\\0\\-1\end{pmatrix}.
\end{equation}
As we will be considering gluon-fusion Higgs production as a pedogogical
example, we will have $Q=m_H$.  From now on the subscript 0 for
jettiness will be implicit. 

The hardness measures $Q_a$ and $Q_b$ in Eq.~\eqref{eq:bigtau} are
arbitrary. The leading-power cross section is not affected by this choice, but the
sub-leading power corrections do depend on these normalization factors.  While our derivation will be valid for
arbitrary $Q_i$, we will call particular attention to two possible
choices: hadronic $\Tau$ and fixed $\Tau$, the former following the
terminology of Ref.~\cite{Moult:2017jsg}. Fixed $\Tau$ is defined by
\begin{equation}
\underline{\text{Fixed:}}\;\;\;Q_a = Q_b = Q, \qquad \qquad \Tau_\text{fix} = \min  \left\{  \frac{2 q_a\cdot p_3}{Q}~,~  \frac{2 q_b \cdot p_3}{Q} \right\},
\end{equation}
while hadronic $\Tau$ is defined by
\begin{equation}
\underline{\text{Hadronic:}}\;\;\;Q_a = x_a \sqrt{s}, \quad Q_b = x_b \sqrt{s}, \qquad \qquad \Tau_\text{had} = \min \left\{ n \cdot p_3, \overline{n} \cdot p_3\right\}.
\end{equation}
%

\section{Leading-power derivation using SCET}
\label{sec:SCETresult}

As a check on our direct QCD derivation of the 0-jettiness
power corrections in Section~\ref{sec:qcd} we
will compare the leading-power expression obtained there with the
result from the SCET factorization
theorem~\cite{Stewart:2010tn,Berger:2010xi}.  According to this result
the differential cross section in $\Tau$ to leading order in the $\Tau
/Q$ expansion can be expressed in terms of
universal objects in the effective theory known as hard, soft and beam
functions that respectively describe hard radiation, soft radiation,
and radiation collinear to a beam direction.  We can write the
leading-power (LP) result as~\cite{Stewart:2010tn,Berger:2010xi}
\begin{align}
\label{eq:facform}
\frac{\d \sigma^{\text{LP}}}{\d \Tau} =& \int \d x_a \int \d x_b \int \d \Phi_H
\int \d \Tau_a \,\d \Tau_b \, 
\delta(\Tau-\Tau_a-\Tau_b)\, \int \d t_a \, \d t_b \, B_a(t_a,x_a,\mu) \,B_b(t_b,x_b,\mu) \,
\nonumber \\ \times & S\left(\Tau_a - \frac{t_a}{Q_a},\Tau_b - \frac{t_b}{Q_b},\mu\right)\,\,H\left(p_H,\mu \right).
\end{align}
Here, $\Phi_H$ denotes the Born phase space for Higgs production,
$x_a$ and $x_b$ denote the Bjorken-$x$ variables for each beam, and
$H$, $B$, and $S$ respectively denote the hard, beam and soft
functions.  The variables $t_a$ and $t_b$ parameterize the
contributions of the beam sectors to the total 0-jettiness
$\Tau$. For gluon-fusion Higgs production the Born-level phase space takes the form 
\begin{equation}
\int \d \Phi_H = \frac{(2 \pi)}{2 s x_a x_b} \int \d^d p_H \, \, \delta(p_H^2 - m_H^2)\delta^{(d)} (q_a+q_b-p_H) =\frac{(2 \pi)}{2 s x_a x_b} \delta (s x_a x_b - m_H^2).
\end{equation}

The factorization formula of Eq.~(\ref{eq:facform}) is true to all
orders in the strong coupling constant $\alpha_s$.  In our study we are interested in the leading-power result
at NLO, so we can expand each function to the NLO level to obtain the fixed-order
result for the differential cross section. Using a convolution symbol
to abbreviate the integrals appearing in Eq.~(\ref{eq:facform}), we
find the following four convolutions contributing to the NLO cross section:
\begin{align}
\frac{\d \sigma_{\text{NLO}}^{\text{LP}}}{\d \Tau}  = &  B_a^{(1)}\otimes B_b^{(0)}\otimes S^{(0)} \otimes H^{(0)}
 ~+~B_a^{(0)}\otimes B_b^{(1)}\otimes S^{(0)} \otimes H^{(0)}\nonumber \\
+~&B_a^{(0)}\otimes B_b^{(0)}\otimes S^{(1)} \otimes H^{(0)}~ +~
 B_a^{(0)}\otimes B_b^{(0)}\otimes S^{(0)} \otimes H^{(1)}.
 \label{eq:convolution}
\end{align}
We have introduced a superscript on each function to denote the order
of $\alpha_s$ contributing to the convolution (for example, $S^{(0)}$
denotes the leading-order soft function, while $S^{(1)}$ denotes the
${\cal O}(\alpha_s)$ coefficient of the soft function).

We will now use the known results from the literature to separately
derive the contributions above.  We begin by compiling the various
functions required in the factorization theorem. For simplicitly we
focus in this section on those terms that involve the gluon
distribution function.  Contributions involving the quark distribution
functions can be obtained in an identical fashion.
\begin{itemize}

\item The beam function is a non-perturbative object that can be
  written as a convolution of perturbative matching coefficients with
  the usual parton distribution functions:
\begin{equation}
\label{eq:beamdef}
B_a \left(t_a, x_a, \mu \right) = \sum_i \int_{x_a}^1 \frac{\d z_a}{z_a} \, \mathcal{I}_{ai} \left( t_a, z_a,\mu\right) f_i \left(\frac{x_a}{z_a},\mu\right).
\end{equation}
The matching coefficients have an expansion in $\alpha_s$:
\begin{equation}
\mathcal{I}_{ij} =\sum_{n=0}^{\oo} \left(\frac{\alpha_s}{4 \pi}\right)^{n} \mathcal{I}_{ij}^{(n)}.
\end{equation}
A similar expansion for the beam function can be obtained upon
substituting the matching coefficients into Eq.~(\ref{eq:beamdef}):
\begin{equation}
B_a(t,x_a,\mu) =\sum_{n=0}^{\oo} \left(\frac{\alpha_s}{4 \pi}\right)^{n} B^{(n)}_a(t,x_a,\mu) .
\end{equation}
The matching coefficients through NNLO can be found
  in~\cite{Gaunt:2014cfa}.  For our purposes we need only the LO and
  NLO results.  Simplifying the relevant expressions from this work
  and keeping only those terms containing the gluon PDF for
  simplicity, we find
\begin{align}
B_g^{(0)} \left(t_a, x_a, \mu \right) =&  \delta(t_a) \, f_g
\left(x_a\right), \nonumber \\
B_g^{(1)} \left(t_a, x_a, \mu \right)=& (4 C_A)\int_{x_a}^1 \frac{\d z_a}{z_a } f_g\left( \frac{x_a}{z_a }\right) \Bigg\{\frac{1}{\mu^2} \mathcal{L}_1 \left( \frac{t_a}{\mu^2}\right) \delta(1-z_a) \nonumber \\
&\hspace{-1.0cm} +\frac{1}{\mu^2} \mathcal{L}_0 \left( \frac{t_a}{\mu^2}\right) \mathcal{L}_0(1-z_a)\frac{\left( 1-z_a + z_a^2\right)^2}{z_a}\nonumber \\
&\hspace{-1.0cm} +\frac{\delta \left( t_a\right)}{2 Q_a} \left[ 2\left(\mathcal{L}_1(1-z_a) -\log z_a \mathcal{L}_0(1-z_a)\right)\frac{\left( 1-z_a+z_a^2\right)^2}{z_a} - \frac{\pi^2}{6}\delta(1-z_a)\right]\Bigg\}.
\end{align}

\item The soft function for 0-jettiness at the NLO level can be obtained
from Ref.~\cite{Jouttenus:2011wh}.  The tree-level result is
\begin{equation}
S^{(0)} =\delta\left(\mathcal{T}_a - \frac{t_a}{Q_a}\right) \delta\left(\mathcal{T}_b - \frac{t_b}{Q_b}\right),
\end{equation}
while the one loop correction can be parameterized as
\begin{equation}
S^{(1)}= \sum_{i\neq j}\mathbf{T}_i \cdot \mathbf{T}_j \, \, S_{ij}^{(1)}.
\end{equation}
In our case, the indices $i$ and $j$ can only take the values
$1,2$. We have
\begin{equation}
\sum_{i\neq j}\mathbf{T}_i \cdot \mathbf{T}_j = - C_A.
\end{equation}
Furthermore, we only have the hemisphere contribution to the soft
function in the language of Ref.~\cite{Jouttenus:2011wh}:
\begin{equation}
S^{(1)}_{ab}= \frac{\alpha_s(\mu)}{4 \pi} \left[\frac{8}{\sqrt{\hat{s}_{ab}}\mu} \mathcal{L}_1\left(\frac{\mathcal{T}_a - \frac{t_a}{Q_a}}{\sqrt{\hat{s}_{ab}} \mu}\right)- \frac{\pi^2}{6} \delta \left(\mathcal{T}_a - \frac{t_a}{Q_a} \right)\right] \delta\left(\mathcal{T}_b-\frac{t_b}{Q_b}\right).
\end{equation}
The constant $\hat{s}_{ab}$ is 
\begin{equation}
\hat{s}_{ab}= 2 \hat{q}_a \cdot \hat{q}_b =\frac{s x_a x_b}{Q_a Q_b} = 1,
\end{equation}
where the $\hat{q}_i$ are defined as
\begin{equation}
\hat{q}_i^\mu = \frac{q_i^\mu}{Q_i}. 
\end{equation}
We have assumed for simplicity that the normalization constants are
chosen so that $Q_a Q_b = s x_a x_b$, as is the case for the leptonic
and hadronic definitions of $\Tau$. There is an analogous contribution
$S_{ba}^{(1)}$ where $a\leftrightarrow b$.

\item After renormalization the NLO hard function contains the finite
  contributions to the virtual corrections. We can obtain the
  leading-order hard
  function for Higgs production from~\cite{Dawson:1990zj}
  and~\cite{Djouadi:1991tka}.  The NLO result can be taken from Ref.~\cite{Ahrens:2008nc}.  The tree-level hard function is simply
  the tree level amplitude squared, which in the effective theory with
  $m_H \ll m_{top}$ takes the form
\begin{equation}
H^{(0)}(p_H,\mu) = |\mathcal{M}_\text{Born}|^2= \frac{\alpha_s^2 m_H^2 \hat{s}}{576 v^2 \pi^2} ,
\end{equation}
where $\hat{s} \equiv x_a x_b s$. At tree level, the $d=4$ dimensional version of the Born matrix element is sufficient. At NLO, we have, in the $\overline{\text{MS}}$ scheme,
\begin{equation}
H^{(1)}(p_H,\mu) =H^{(0)}(p_H,\mu) \left(\frac{\alpha_s C_A}{\pi} \right) \left\{ \frac{7}{12} \pi^2 - \frac{1}{2} \log^2 \left(\frac{m_H^2}{\mu^2} \right) \right\}.
\end{equation}
There is as additional correction in the effective-theory Lagrangian
coming from integrating out the top quark.  As this term is treated identically in
SCET and in the direct QCD derivation of the differential cross
section we do not discuss it explicitly here.  It is included in all
numerical results.

\end{itemize}

We must now use these expressions in the expanded factorization
formula of Eq.~(\ref{eq:convolution}).  At the NLO level there is no
non-trivial convolution to perform, since in each case all but one
function takes on its simple tree-level form, and all integrals can be done
straightforwardly.  We separately present the individual contributions
to the differential cross section in $\Tau$ for the gluon-gluon
partonic channel below:
\begin{align}
B_g^{(1)}\otimes B_g^{(0)}\otimes S^{(0)} \otimes H^{(0)}= &\left(\frac{\alpha_s C_A}{ \pi} \right)\int \d x_a \int \d x_b \int \d \Phi_H |\mathcal{M}_\text{Born}|^2 f_g \left(x_b\right) \int_{x_a}^1 \frac{\d z_a}{z_a } f_g\left( \frac{x_a}{z_a }\right)\nonumber \\
& \hspace{-2.0cm} \Bigg\{\frac{Q_a}{\mu^2} \mathcal{L}_1 \left( \frac{Q_a\Tau}{\mu^2}\right) \delta(1-z_a) +\frac{Q_a}{\mu^2} \mathcal{L}_0 \left( \frac{Q_a\Tau}{\mu^2}\right) \mathcal{L}_0(1-z_a)\frac{\left( 1-z_a + z_a^2\right)^2}{z_a}\nonumber \\
&\hspace{-2.0cm}+\delta (\Tau) \left[ \left(\mathcal{L}_1(1-z_a) -\log z_a \mathcal{L}_0(1-z_a)\right)\frac{\left( 1-z_a+z_a^2\right)^2}{z_a} - \frac{\pi^2}{12}\delta(1-z_a)\right]\Bigg\};
\label{eq:beamaNLO}
\end{align}
\begin{align}
B_g^{(0)}\otimes B_g^{(0)}\otimes S^{(1)} \otimes H^{(0)} =&\left( \frac{\alpha_s C_A}{ \pi}\right)\int \d x_a \int \d x_b \int \d \Phi_H |\mathcal{M}_\text{Born}|^2 f_g(x_a) f_g(x_b)\nonumber \\
& \Bigg\{-\frac{4}{ \mu} \mathcal{L}_1\left(\frac{\mathcal{T}}{  \mu}\right)+ \frac{\pi^2}{12} \delta \left(\mathcal{T}\right)\Bigg\};
\label{eq:softNLO}
\end{align}
\begin{align}
B_g^{(0)}\otimes B_g^{(0)}\otimes S^{(0)} \otimes H^{(1)} =&\left( \frac{\alpha_s C_A}{ \pi}\right)\int \d x_a \int \d x_b \int \d \Phi_H |\mathcal{M}_\text{Born}|^2 f_g(x_a) f_g(x_b)\nonumber \\
& \delta(\Tau)\Bigg[ \frac{7}{12} \pi^2 - \frac{1}{2} \log^2 \left(\frac{m_H^2}{\mu^2} \right)\Bigg].
\label{eq:hardNLO}
\end{align}
There is a fourth contribution $B_g^{(0)}\otimes B_g^{(1)}\otimes S^{(0)}
\otimes H^{(0)}$ which can be obtained from Eq.~(\ref{eq:beamaNLO}) with the
substitution $a \leftrightarrow b$.

\section{Direct QCD derivation of leading and sub-leading power}
\label{sec:qcd}

In order to study in detail the structure of the power corrections to
the 0-jettiness factorization formula it is useful to expand the cross
section directly in QCD.  It turns out to be possible
to obtain the full ${\cal O}(\Tau^\text{cut})$ power corrections, not
just the logarithmically-enhanced terms studied previously.  We will
derive here the leading-power result as well to compare to the SCET
expression of the previous section.  This serves as a check on our
result, and we hope that it is also useful to the reader interested in
understanding the connection between the SCET formalism and standard
QCD.  For clarity, the $\Tau$ behavior of the coefficients at
leading-power and next-to-leading-power is shown in
Table~\ref{tab:LPNLP}.  We show the power counting both for the
differential cross section in $\Tau$ and for the result integrated up
to a cutoff $\Tauc$.  We begin by discussing the factorization of the
phase space and the expansion of the matrix elements in $\Tau$.  Most
aspects of our derivation are applicable to arbitrary 0-jettiness
processes and not just gluon-fusion Higgs production.  We identify in the
text which parts of our derivation are generic and which parts must be
modified for other processes.

\begin{table}[h]
\label{tab:LPNLP}
\caption{Power counting in $\Tau$ of the NLO cross section
  differential in $\Tau$ (left) and integrated up to $\Tauc$ (right),
  at leading-power, next-to-leading-power and next-to-next-to-leading
  power.  We note that there is no NNLL contribution at NLO in $\alpha_s$.}
\centering
\vspace{5mm}
\begin{tabular}[c]{lccc}
\toprule
$  \frac{\d \sigma_\text{NLO}}{\d \Tau} $  &LL &NLL &NNLL \\
\colrule
\vspace{-4mm}\\
LP    & ~~$\displaystyle\left[\frac{\log \Tau}{\Tau} \right]_+$~~      & ~$\displaystyle\left[\frac{1}{\Tau} \right]_+$~      & ~~~$\delta \left(\Tau \right)$~~~        \\
\vspace{-4mm}\\
NLP  & $\log \Tau$    & 1     &           \\
\botrule
\end{tabular}
\hspace{1.5cm}
\begin{tabular}[c]{lccc}
\toprule
$    \sigma_\text{NLO} $  &LL &NLL &~~NNLL~~ \\
\colrule
\vspace{-4mm}\\
 LP    & $\displaystyle \log^2 \Tauc$      & $\log \Tauc$      & 1        \\
\vspace{-4mm}\\
NLP  & ~~$\Tauc\log \Tauc$~~    & $\Tauc$    &           \\
\botrule
\end{tabular}
\end{table}
%

\subsection{Factorization of the phase space}

The SCET result for the below-cut region in Eq.~(\ref{eq:facform}) is written in a form that
explicitly factors out the Born-level phase space.  This is possible
directly in QCD with straightforward changes of variables.  The strategy used
here is to absorb the kinematics of the emitted final-state gluon into
one of the two incoming gluons. The approach adopted here works as long as we are
inclusive in the Higgs rapidity. We begin by deriving the Born-level
phase space for the leading-order process
$g(q_a)+g(q_b) \to H$, including also the convolution over parton
distribution functions and the flux factor.  The kinematics for the
initial state was described in Section~\ref{sec:taudef}, from which we find
\begin{align}
\label{eq:PSBorn}
\text{PS}_\text{Born} &=(2 \pi) \int_0^1 \d x_a \int_0^1 \d x_b \, \frac{f_g(x_a) \, f_g(x_b)}{2 s x_a x_b}\,  \int \d^d p_H  \delta(p_H^2 - m_H^2)  \,\delta^{(d)}(q_a + q_b - p_H) \nonumber \\
&=(2 \pi) \int_0^1 \d x_a \int_0^1 \d x_b \, \frac{f_g(x_a) \,
  f_g(x_b)}{2 s x_a x_b}\,\,   \delta(s x_a x_b - m_H^2)  .
\end{align}

At NLO we consider the partonic process
$g(q_a^{\prime})+g(q_b^{\prime}) \to H+g(p_3)$.  The same approach is
applicable to the other NLO partonic channels, which are
$q(q_a^{\prime})+g(q_b^{\prime}) \to H+q(p_3)$,
$g(q_a^{\prime})+q(q_b^{\prime}) \to H+g(p_3)$, and
$q(q_a^{\prime})+\bar{q}(q_b^{\prime}) \to H+g(p_3)$.  For simplicity
we only present the derivation of the most complicated $gg$ case
here.  The sub-leading power results for the other partonic channels are given in the
Appendix.  We have relabeled the
initial-state gluon momenta to distinguish them from the momenta entering the
0-jettiness definition of Eq.~(\ref{eq:bigtau}).  Denoting the
  initial parton momentum fractions $\xi_a$ and $\xi_b$ to distinguish them from
  those entering the 0-jettiness definition in Eq.~(\ref{eq:bigtau}), we have
\begin{align}
\text{PS}_\text{NLO} &=\frac{ \mu_0^{2 \varepsilon}}{\left( 2
    \pi\right)^{d-2}} \int_0^1 \d \xi_a \int_0^1 \d \xi_b \, \frac{f_g(\xi_a) \, f_g(\xi_b)}{2 s \xi_a \xi_b }\int \d^{d} p_3 \,\delta(p_3^2)   \nonumber \\
&\times \int \d^{d} p_H  \delta(p_H^2 - m_H^2) \,\delta^{(d)}(q_a^{\prime} + q_b^{\prime} - p_3
- p_H)\Theta \left(s \xi_a \xi_b- m_H^2  \right) \Theta(E_H)
\nonumber \\
&=\frac{ \mu_0^{2 \varepsilon}}{\left( 2 \pi\right)^{d-2}} \int_0^1 \d \xi_a \int_0^1 \d \xi_b \, \frac{f_g(\xi_a) \, f_g(\xi_b)}{2 s \xi_a \xi_b }\int \d^{d} p_3 \,\delta(p_3^2)    \delta\left(\hat{s} - m_H^2 - 2 q_a^{\prime} \cdot p_3 - 2 q_b^{\prime} \cdot p_3\right).
\end{align}
We have included $\mu_0$, the usual dimensional-regularization mass
scale, in the phase space.   In the second line we have used
momentum conservation to remove the $p_H$ integral, and have
suppressed the energy theta-function for notational simplicity.  We use a Sudakov decomposition for the gluon momentum $p_3$:
\begin{align}
p_3^\mu &= \frac{\overline{n} \cdot p_3}{2} \, n^\mu +  \frac{n \cdot
  p_3}{2} \, \overline{n}^\mu + p_{3T}^\mu, \nonumber \\
\int \d^d p_3 \delta \left(p_3^2\right)&=\frac{\Omega_{d-2}}{4} \int \d (\overline{n} \cdot p_3) \, \, \left(\overline{n} \cdot p_3\right)^{-\varepsilon}\int \d ( n \cdot p_3) \, \,  \left( n \cdot p_3\right)^{-\varepsilon} .
\end{align}

We now incorporate the definition of 0-jettiness into the phase
space.  The emitted gluon $p_3$ can be closer to either the $n$ or
$\overline{n}$ direction, leading to different expressions for $\Tau$
according to the definition in Eq.~(\ref{eq:bigtau}).  We will assume
the first case.  The second case can be easily obtained by exchanging $a
\leftrightarrow b$.  In this region we have
\begin{equation}
\Tau = \frac{x_a \sqrt{s}}{Q_a} n \cdot p_3.
\end{equation}
We note that $\Tau$ has been defined using the $q_a$ and $q_b$ defined
in Eq.~(\ref{eq:qdef}).  We will see how to define these quantities at
NLO shortly.  We can now write down the differential phase space for
this partition as
\begin{align}
\frac{\d\text{PS}^{(a)}_\text{NLO}}{\d \Tau} =\frac{ \Tau^{-\varepsilon}}{8 \pi} \frac{\left( 4 \pi \mu_0^2\right)^\varepsilon}{\Gamma(1-\varepsilon)}&\int_0^1 \d \xi_a \int_0^1 \d \xi_b \frac{f_g(\xi_a) f_g(\xi_b)}{2 s \xi_a \xi_b} \left( \frac{Q_a}{x_a \sqrt{s}}\right)^{1-\varepsilon}\nonumber \\
&\int \d (\overline{n} \cdot p_3) \, \, (\overline{n} \cdot p_3)^{-\varepsilon} \delta\left( s \xi_a \xi_b - m_H^2- Q_a  \frac{\xi_a}{x_a}\Tau - (\overline{n} \cdot p_3) \xi_b \sqrt{s}\right).
\end{align}
We rescale $\overline{n} \cdot p_3$ so that it ranges from 0 to 1
using the variable change
\begin{equation}
\overline{n} \cdot p_3 = \sqrt{s} \xi_a (1-z_a),
\end{equation}
leading to the expression
\begin{align}
\label{eq:diffPSintermed}
\frac{\d\text{PS}^{(a)}_\text{NLO}}{\d \Tau} =\frac{ \Tau^{-\varepsilon}}{8 \pi} \frac{\left( 4 \pi \mu_0^2\right)^\varepsilon}{\Gamma(1-\varepsilon)}&\int_0^1 \d \xi_a \int_0^1 \d \xi_b \frac{f_g(\xi_a) f_g(\xi_b)}{2 s \xi_a \xi_b} \left( \frac{Q_a \xi_a}{x_a }\right)^{1-\varepsilon}\nonumber \\
&\int \d z_a  (1-z_a)^{-\varepsilon} \delta\left( s \xi_a \xi_b z_a - m_H^2- Q_a  \frac{\xi_a}{x_a}\Tau \right).
\end{align}

At this point we wish to identify and factor out the Born phase space
defined in Eq.~(\ref{eq:PSBorn}) from this expression.  We can do so
with the following variable changes, which also serve the purpose of
defining the $x_a$ and $x_b$ that appear in the 0-jettiness definition:
\begin{equation}
\xi_a = \frac{s x_a^2 x_b}{s x_a x_b z_a - Q_a \Tau} \qquad \qquad
\xi_b = x_b .
\end{equation}
These transformations force the delta function in Eq.~(\ref{eq:diffPSintermed}) into the form which
appears in the LO phase space.  We note one point that
appears when we perform this variable change.  In general,  $Q_a$ is 
a function of $x_a$ (for example, for the hadronic $\Tau$
it is $Q_a(x_a) = \sqrt{s} x_a$).   Similarly, $Q_b=Q_a(x_b)$.  Therefore derivatives of the $Q_i$ appear when changing
variables from $\xi_a$ to $x_a$. Upon making these variable changes
and expanding the phase-space measure to ${\cal O}(\Tau)$, we have
\begin{align}
\label{eq:QCDPS}
\frac{\d\text{PS}^{(a)}_\text{NLO}}{\d \Tau} &=\frac{ \Tau^{-\varepsilon}}{8 \pi} \frac{\left( 4 \pi \mu_0^2\right)^\varepsilon}{\Gamma(1-\varepsilon)}\int_0^1 \d x_a \int_0^1 \d x_b \frac{ f_g(x_b)}{2 s  x_a x_b} \delta\left(s x_a x_b - m_H^2\right)\int_{x_a  + \frac{Q_a \Tau}{m_H^2}}^{ \frac{m_H^4 + \Tau^2 Q_a Q_b}{m_H^2 \left( m_H^2 + \Tau Q_b\right)}} \frac{\d z_a }{z_a} \nonumber \\
& \left(  Q_a  \right)^{1-\varepsilon} \left(\frac{1-z_a}{z_a}\right)^{-\varepsilon}\Bigg\{ f_g \left(\frac{x_a}{z_a} \right) + \frac{ \Tau}{m_H^2 z_a^2}\left[ \left(Q'_a z_a x_a- Q_a z_a \varepsilon\right) {f_g}\left( \frac{x_a}{z_a}\right) + Q_a x_a f'_g\left(\frac{x_a}{z_a} \right)\right] \nonumber \\
&+ \mathcal{O} \left( \Tau^2\right)\Bigg\}.
\end{align}
The prime denotes a derivative with respect to
$x_a$. This has the desired factorized Born-level phase space.  
 We note that the lower limit on the $z_a$
integral comes from the requirement $\xi_a \leq 1$, while the upper
limit comes from the 0-jettiness requirement 
\begin{equation}
\Tau \le  \frac{x_b \sqrt{s} \overline{n}\cdot p_3 }{Q_b} \qquad \implies \qquad z_a \le  \frac{m_H^4 + \Tau^2 Q_a Q_b}{m_H^2 \left( m_H^2 + \Tau Q_b\right)} = 1-\frac{Q_b \Tau}{m_H^2} + \frac{Q_b\left( Q_a+Q_b\right)\Tau^2}{m_H^4}+\mathcal{O}\left( \Tau^3\right)
\end{equation}
We have checked that the $z_a$ integral can be extended down to
$x_a$, since the region between $x_a$ and $x_a  + \frac{Q_a
  \Tau}{m_H^2}$ does not contribute to the LL nor to the  NLL power
corrections.  We will express our results in the usual $\overline{\text{MS}}$ scheme
by replacing
\begin{equation}
\left(4 \pi \mu_0^2\right)^\varepsilon = \left(e^{\gamma_E}\mu^2\right)^\varepsilon.
\end{equation}
We note that this factorization of the phase space is valid for any process using
0-jettiness as a resolution parameter.

\subsection{Expansion of the matrix elements}

We now consider the matrix elements for the NLO real-emission correction $gg \to Hg$, expanded in
$\Tau$.  This partonic channel is numerically the most important
contribution to gluon-fusion Higgs production. 
Results for these other channels are given in the Appendix.   We note that the structure of the power corrections takes the form of
leading-power phase space combined with sub-leading power matrix
elements, plus sub-leading power phase space combined with
leading-power matrix elements.  While the exact expressions for the
matrix elements are of course process-dependent,
the structure of the power corrections is the same for any 0-jettiness
process. 

The Born matrix element for the partonic process $gg \to H$ in $d$ dimensions is~\cite{Dawson:1990zj}
\begin{equation}
\left|\mathcal{M}(gg \to H)\right|^2 = \frac{\alpha_s^2 m_H^4}{576 \pi^2 v^2}\left( \frac{\mu^2}{m_t^2}\right)^{2\varepsilon} \frac{\Gamma^2(1+\varepsilon)}{1-\varepsilon}.
\end{equation}
For the real emission of a gluon, we have 
\begin{align}
|\mathcal{M}(gg \to Hg)|^2=  \frac{8 C_A \pi \alpha_s}{m_H^4(1-\varepsilon)}\left|\mathcal{M}(gg\to H) \right|^2&\Bigg[\frac{m_H^8 + s_{12}^4 + s_{13}^4+s_{23}^4 }{s_{12} s_{13} s_{23}}(1-2\varepsilon) \nonumber \\
&+ \frac{\varepsilon}{2}\frac{\left(m_H^4+ s_{12}^2 + s_{13}^2+s_{23}^2 \right)^2}{s_{12} s_{13} s_{23}}\Bigg].
\label{eq:NLOME}
\end{align}
The invariants that appear in the matrix elements are given in our
phase-space parameterization by
\begin{equation}
s_{12} = s \xi_a \xi_b = \frac{m_H^2}{z_a - \frac{Qa \Tau}{m_H^2}},\;\;
s_{13} = - \frac{Q_a \Tau}{z_a - \frac{Qa \Tau}{m_H^2}},\;\;
s_{23} =  - m_H^2 \frac{1-z_a}{z_a - \frac{Qa \Tau}{m_H^2}}.
\end{equation}
Therefore, we can write the NLO matrix element including both the
leading power in ${\cal T}$ and the first correction as
\begin{align}
\label{eq:QCDME}
|\mathcal{M}(gg \to Hg)|^2 &=  |\mathcal{M}(gg \to H)|^2 \left( 16 C_A \alpha_s \pi\right) \Bigg\{\frac{1}{Q_a \Tau} \frac{\left( 1-z_a+z_a^2\right)^2}{(1-z_a)z_a} \nonumber \\
&+ \frac{1}{m_H^2} \left[5 -\frac{1}{1-z_a} +\frac{1}{z_a^2} - \frac{1}{z_a} + z_a -\frac{2}{1-\varepsilon} \right]+ \mathcal{O}(\Tau)\Bigg\}.
\end{align}

\subsection{Derivation of the leading-power result}

We begin by deriving the leading-power expression to compare with our
result of Section~\ref{sec:SCETresult}.  To obtain this term we take the
leading-power expression for both the phase space and the matrix
element.  We can write the differential cross section for the first
phase-space partition as
\begin{align}
\frac{\d\sigma^{\text{LP},(a)}_\text{NLO}}{\d \Tau} =&  \left(\frac{C_A \alpha_s}{\pi} \right)\frac{\left( e^{\gamma_E}\right)^\varepsilon}{\Gamma(1-\varepsilon)}\int_0^1 \d x_a \int_0^1 \d x_b \frac{ (2 \pi) f_g(x_b)}{2 s  x_a x_b} |\mathcal{M}(gg \to H)|^2\delta\left(s x_a x_b - m_H^2\right)\nonumber \\
&\left(\frac{\Tau Q_a}{\mu^2}\right)^{-1-\varepsilon} \frac{Q_a}{\mu^2}   \int_{x_a  }^{\frac{m_H^4 + \Tau^2 Q_a Q_b}{m_H^2 \left( m_H^2 + \Tau Q_b\right)} }\frac{\d z_a }{z_a} f_g \left(\frac{x_a}{z_a} \right) \left( 1-z_a\right)^{-1-\varepsilon} z_a^\varepsilon  \frac{\left( 1-z_a+z_a^2\right)^2}{z_a} .
\end{align}
To proceed we divide the integral in $z_a$ into two regions as
follows:
\begin{equation}
\label{eq:zsplit}
 \int_{x_a  }^{ \frac{m_H^4 + \Tau^2 Q_a Q_b}{m_H^2 \left( m_H^2 + \Tau Q_b\right)} } \d z_a = \int_{x_a  }^{ 1} \d z_a- \int_{\frac{m_H^4 + \Tau^2 Q_a Q_b}{m_H^2 \left( m_H^2 + \Tau Q_b\right)}  }^{ 1} \d z_a.
\end{equation}
We can now establish the connection between the direct QCD calculation
performed here and the SCET result of Section~\ref{sec:SCETresult}.  
The first integral will give us the beam function contribution of
SCET, while the second integral will give us the the soft
function contribution.  We expand in $\varepsilon$ using
\begin{equation}
 \frac{Q_a}{\mu^2}\left( \frac{Q_a \Tau}{\mu^2}\right)^{-1-\varepsilon} ~\to~ -\frac{\delta(\Tau) }{\varepsilon} +  \frac{Q_a}{\mu^2}\mathcal{L}_0 \left( \frac{Q_a \Tau}{\mu^2}\right) - \varepsilon  \frac{Q_a}{\mu^2}\mathcal{L}_1 \left( \frac{Q_a \Tau}{\mu^2}\right),
\end{equation}
\begin{equation}
(1-z_a)^{-1-\varepsilon} z_a^\varepsilon~\to~ -\frac{\delta(1-z_a)}{\varepsilon} + \mathcal{L}_0(1-z_a) - \varepsilon \left[\mathcal{L}_1(1-z_a) - \log z_a \, \mathcal{L}_0 (1-z_a)\right],
\end{equation}
where the plus distribution is written as
\begin{equation}
\mathcal{L}_n(x) = \left[ \frac{\theta(x)\log^nx}{x}\right]_+ .
\end{equation}
Using these expressions the differential cross section can be written
as
\begin{align}
\frac{\d\sigma^{\text{LP},(a)}_\text{beam}}{\d \Tau} &=\left( \frac{\alpha_s C_A}{\pi}\right)\int_0^1 \d x_a \int_0^1 \d x_b (2 \pi) \frac{ f_g(x_b)}{2 s x_a x_b}|\mathcal{M}_\text{Born}|^2 \delta\left( s x_a x_b - m_H^2\right)\int_{x_a}^1\frac{\d z_a}{z_a}f_g \left( \frac{x_a}{z_a}\right)\nonumber \\
& \Bigg\{ \frac{1}{\varepsilon^2} \delta\left(\Tau\right) \delta(1-z_a)- \frac{1}{\varepsilon} \mathcal{L}_0 (1-z_a) \frac{\left( 1-z_a+z_a^2\right)^2}{ z_a}\delta(\Tau)- \frac{1}{\varepsilon} \left(\frac{Q_a}{\mu^2} \right) \mathcal{L}_0\left(\frac{Q_a \Tau}{\mu^2} \right) \delta(1-z_a) \nonumber \\
&+ \left(\frac{Q_a}{\mu^2} \right) \mathcal{L}_1\left(\frac{Q_a \Tau}{\mu^2} \right) \delta(1-z_a) +\left( \frac{Q_a}{\mu^2}\right) \mathcal{L}_0 \left( \frac{Q_a \Tau}{\mu^2}\right)\mathcal{L}_0 (1-z_a)\frac{\left( 1-z_a+z_a^2\right)^2}{ z_a}\nonumber \\
&- \frac{\pi^2}{12} \delta(\Tau)\delta(1-z_a)+\left[\mathcal{L}_1(1-z_a) - \log z_a \, \mathcal{L}_0 (1-z_a)\right]\frac{\left( 1-z_a+z_a^2\right)^2}{ z_a}\delta(\Tau)\Bigg\}.
\end{align}
The term inside the braces, together with the overall factor outside
of the integral, is exactly the bare gluon beam
function~\cite{Berger:2010xi}. The finite terms are the same that we
found in Eq.~(\ref{eq:beamaNLO}) by expanding the SCET
factorization formula.  In the effective theory the poles would be
removed by a separate renormalization of the beam function and by
matching to the PDFs.  In our
QCD calculation they cancel upon adding all contributions to the cross
section, including the mass factorization counterterms.

We now consider the second region of the $z_a$ integral in Eq.~(\ref{eq:zsplit}).
We can expand the integrand around $z_a=1$:
\begin{equation}
\int_{\frac{m_H^4 + \Tau^2 Q_a Q_b}{m_H^2 \left( m_H^2 + \Tau Q_b\right)}}^1 \d z_a \Bigg\{(1-z_a)^{-1-\varepsilon} f(x_a) + (1-z_a)^{-\varepsilon}\left[\varepsilon f(x_a) - x_a f'_g(x_a) \right] \Bigg\}.
\label{eq:zexpansion}
\end{equation}
The first term in this expansion contributes to the leading-power and to the sub-leading power result\footnote{The sub-leading correction coming from this term was found thanks to discussions with the authors of~\cite{Ebert:2018lzn}.}.  The second term contributes only to sub-leading power, and
will be needed in the next sub-section.  For the first term, the following integral is needed:
\begin{equation}
\int_{\frac{m_H^4 + \Tau^2 Q_a Q_b}{m_H^2 \left( m_H^2 + \Tau Q_b\right)}}^1 \d z_a (1-z_a)^{-1-\varepsilon}= \left( \frac{Q_b \Tau }{m_H^2}\right)^{-\varepsilon} \left[ - \frac{1}{\varepsilon} - \frac{\left( Q_a + Q_b\right) \Tau}{m_H^2}\right].
\label{eq:intz1}
\end{equation}
We compile here for use in the sub-leading power derivation the
following result as well: 
\begin{equation}
\int_{\frac{m_H^4 + \Tau^2 Q_a Q_b}{m_H^2 \left( m_H^2 + \Tau Q_b\right)}}^1 \d z_a (1-z_a)^{-\varepsilon}= \frac{\left( \frac{Q_b \Tau}{m_H^2}\right)^{1-\varepsilon}}{1-\varepsilon}.
\label{eq:intz2}
\end{equation}
We again assume for simplicity that $Q_a Q_b = m_H^2$ as we did in
Section~\ref{sec:SCETresult}, leading to the result:
\begin{align}
\frac{\d\sigma^{\text{LP},(a)}_\text{soft}}{\d \Tau} &=\left( \frac{\alpha_s C_A}{\pi}\right)\int_0^1 \d x_a \int_0^1 \d x_b (2\pi)\frac{f_g(x_a) f_g(x_b)}{2 s x_a x_b} |\mathcal{M}_\text{Born}|^2\nonumber \\
&\Bigg\{ - \frac{1}{2 \varepsilon^2} \delta (\Tau) + \frac{1}{\varepsilon \mu} \mathcal{L}_0 \left( \frac{\Tau}{\mu}\right) - \frac{2}{\mu} \mathcal{L}_1 \left( \frac{\Tau}{\mu}\right) + \frac{\pi^2}{24} \delta (\Tau)\Bigg\}.
\end{align}
We can again compare the finte part of this result to the SCET soft-function
contribution of Eq.~(\ref{eq:softNLO}).  This is equal to half of the
result there.   The remaining factor of two is provided by the other
phase-space partition, establishing the exact correspondence between the
QCD calculation and the SCET derivation.  We have performed this check
for the other partonic channels as well.

It only remains to consider the virtual corrections to the cross
section and to establish the cancellation of poles.  Since the hard
function of SCET is exactly the finite part of the virtual corrections
in QCD as can be checked by comparing Refs.~\cite{Ahrens:2008nc}
and~\cite{Harlander:2000mg}, the correspondence between the QCD
virtual corrections and the SCET hard-function contribution of
Eq.~(\ref{eq:hardNLO}) is obvious.  

We now focus on the poles, since we have already established the agreement
of the finite parts between SCET and QCD for the separate
contributions.  We can combine both beam regions
and both soft contributions to get the full result for the
real-emission correction at leading power:
\begin{equation}
\frac{\d\sigma^{\text{LP}}_{\text{real}}}{\d \Tau} = \frac{\d\sigma^{\text{LP} ,(a)}_\text{beam}}{\d \Tau} +\frac{\d\sigma^{\text{LP} ,(a)}_\text{soft}}{\d \Tau} +\frac{\d\sigma^{\text{LP} (b)}_\text{beam}}{\d \Tau} +\frac{\d\sigma^{\text{LP} ,(b)}_\text{soft}}{\d \Tau}. 
\end{equation}
Combining these terms we arrive at
\begin{align}
\frac{\d\sigma^{\text{LP}}_{\text{real}}}{\d \Tau} =\left( \frac{\alpha_s C_A}{\pi}\right)&\int_0^1 \d x_a \int_0^1 \d x_b \frac{(2\pi)|\mathcal{M}_\text{Born}|^2}{2 s x_a x_b} \delta\left( s x_a x_b - m_H^2\right)\int_{x_a}^1 \frac{\d z_a}{z_a}f_g\left( \frac{x_a}{z_a}\right)\int_{x_b}^1 \frac{\d z_b}{z_b}f_g\left( \frac{x_a}{z_b}\right)\nonumber \\
&\delta\left(\Tau\right)\Bigg\{\delta(1-z_a) \delta(1-z_b)\left[ \frac{1}{\varepsilon^2}   -\frac{1}{\varepsilon} \log \left(\frac{m_H^2}{\mu^2} \right) \right]   \nonumber \\ &- \frac{1}{\varepsilon} \mathcal{L}_0 (1-z_a) \frac{\left( 1-z_a+z_a^2\right)^2}{z_a}- \frac{1}{\varepsilon} \mathcal{L}_0 (1-z_b)\frac{\left( 1-z_b+z_b^2\right)^2}{z_b} \Bigg\}.
\end{align}
We note that in writing this expression we have made use of the
following relation:
\begin{equation}
\begin{split}
 2\mathcal{L}_0\left( \frac{\Tau}{\mu}\right)-\left(\frac{Q_a}{\mu^2}
 \right)\mathcal{L}_0\left(\frac{Q_a \Tau}{\mu^2}
 \right)-\left(\frac{Q_b}{\mu^2} \right)\mathcal{L}_0\left(\frac{Q_b
     \Tau}{\mu^2} \right)&=-\frac{1}{\varepsilon} \log \left(\frac{Q_a
     Q_b}{\mu^2} \right) \delta(\Tau) \\ &=-\frac{1}{\varepsilon} \log \left(\frac{m_H^2}{\mu^2} \right) \delta(\Tau) .
\end{split}
\end{equation}
This is the expression for the poles of the real-emission
corrections.  The first two terms cancel against the virtual
corrections, whose pole structure at NLO  in the
$\overline{\text{MS}}$ scheme is given by~\cite{Harlander:2000mg}
\begin{equation}
\sigma_V = \sigma_\text{Born} \left( \frac{\alpha_s C_A}{\pi}\right) \left(\frac{\mu^2}{m_H^2} \right)^\varepsilon \left[ -\frac{1}{\varepsilon^2}\right] = \sigma_\text{Born} \left( \frac{\alpha_s C_A}{\pi}\right)  \left[ -\frac{1}{\varepsilon^2}+\frac{1}{\varepsilon} \log \left( \frac{m_H^2}{\mu^2}\right)\right].
\end{equation}
The second two terms are removed by mass factorization into the
initial-state PDFs, leaving a finite cross section.  This completely
establishes both the cancellation of poles and the equivalence of QCD
and SCET at leading power in $\Tau$.  We note that we have not
considered the additional contribution in the Higgs effective theory
coming from integrating out the top quark, since it is treated identically
in both a direct QCD calculation and in the effective theory.

\subsection{Derivation of the subleading-power result}

Having established that our direct QCD calculation reproduces the leading-power result of the
SCET factorization theorem, we proceed to study the
next-to-leading-power (NLP) using the same approach.  There are three
sources of corrections to consider: sub-leading power terms in
the phase space expression of Eq.~(\ref{eq:QCDPS}), the sub-leading power
correction in the matrix element in Eq.~(\ref{eq:QCDME}), and the
expansion of the soft-region integral of Eq.~(\ref{eq:zexpansion}) to sub-leading
order. The contributions from the first two pieces can be
divided into beam and soft regions, following the split of the $z_a$
integral performed for the leading-power term in
Eq.~(\ref{eq:zsplit}).  The beam-region contribution to the
NLP result takes the form
\begin{align}
\frac{\d\sigma^{\text{NLP},(a)}_\text{beam}}{\d \Tau} =& \left( \frac{C_A \alpha_s}{\pi}\right)\int_0^1 \d x_a \int_0^1 \d x_b \frac{(2 \pi) f_g(x_b)}{2 s  x_a x_b} \delta\left(s x_a x_b - m_H^2\right) |\mathcal{M}(gg \to H)|^2 \nonumber \\
& \frac{Q_a}{m_H^2}\int_{x_a  }^{ 1} \frac{\d z_a }{z_a}  \Bigg\{   \frac{\delta(1-z_a)}{\varepsilon}  \left[\left(1-x_a\frac{Q'_a}{Q_a}\right)f_g \left( x_a\right) - x_a f'_g(x_a) \right]\nonumber  \\
&+\delta(1-z_a) \log \left( \frac{Q_a \Tau}{\mu^2}\right)\left[ \left(-1+\frac{Q'_a}{Q_a} x_a\right)f_g(x_a)+x_a f'_g(x_a)\right] \nonumber \\
&-f_g \left(\frac{x_a}{z_a} \right)\mathcal{L}_0 (1-z_a)+  f_g(x_a) \delta(1-z_a) +f_g \left(\frac{x_a}{z_a} \right)  \left[3  +\frac{1}{z_a^2} - \frac{1}{z_a} + z_a  \right]\nonumber \\
&+ x_a\frac{\left( 1-z_a+z_a^2\right)^2}{z_a^3}\mathcal{L}_0(1-z_a) \left[ f_g'\left(\frac{x_a}{z_a} \right) +\frac{Q'_a}{Q_a}z_a f_g\left( \frac{x_a}{z_a}\right)\right]\Bigg\}.
\end{align}
For the soft region we also must include the contribution from the
second term in the right hand side of~\eqref{eq:intz1} and from the term in the right hand side of~\eqref{eq:intz2}, leading to the total soft contribution
\begin{align}
\frac{\d\sigma^{\text{NLP},(a)}_\text{soft}}{\d \Tau} =& \left( \frac{C_A \alpha_s}{\pi}\right)\int_0^1 \d x_a \int_0^1 \d x_b \frac{(2 \pi) f_g(x_b)}{2 s  x_a x_b} \delta\left(s x_a x_b - m_H^2\right) |\mathcal{M}(gg \to H)|^2 \nonumber \\
&\frac{Q_a}{m_H^2}\Bigg\{ -\frac{1}{\varepsilon} \left[\left( 1-\frac{Q'_a x_a}{Q_a}\right)f_g(x_a) -x_a f'_g(x_a) \right]+\frac{Q_b}{Q_a} f_g(x_a)\nonumber \\
& +\log\left( \frac{\Tau^2}{\mu^2}\right)\left[\left( 1-\frac{Q'_a x_a}{Q_a}\right)f_g(x_a) - x_a f'_g(x_a) \right]-\left( \frac{Q_b }{Q_a}\right)   x_a f'_g(x_a)\Bigg\} .
\end{align}
The full NLP correction when the emitted gluon is close to the $n^\mu$ direction can be found by simply summing the beam and soft contribution. As expected, the apparent pole cancels:
\begin{align}
\label{eq:fullPC1}
\frac{\d\sigma^{\text{NLP},(a)}}{\d \Tau} =& \left( \frac{C_A \alpha_s}{\pi}\right)\int_0^1 \d x_a \int_0^1 \d x_b \frac{(2 \pi) f_g(x_b)}{2 s  x_a x_b} \delta\left(s x_a x_b - m_H^2\right) |\mathcal{M}(gg \to H)|^2 \nonumber \\
& \hspace{-1.0cm} \frac{Q_a}{m_H^2}\int_{x_a  }^{ 1} \frac{\d z_a }{z_a}  \Bigg\{  -f_g \left(\frac{x_a}{z_a} \right)\mathcal{L}_0 (1-z_a)+f_g \left(\frac{x_a}{z_a} \right)  \left[3  +\frac{1}{z_a^2} - \frac{1}{z_a} + z_a  \right]\nonumber  \\
& \hspace{-1.0cm} + x_a\frac{\left( 1-z_a+z_a^2\right)^2}{z_a^3}\mathcal{L}_0(1-z_a) \left[ f_g'\left(\frac{x_a}{z_a} \right) +\frac{Q'_a}{Q_a}z_a f_g\left( \frac{x_a}{z_a}\right)\right] + \delta(1-z_a) f_g(x_a)\nonumber \\
& \hspace{-1.0cm}+\delta(1-z_a)\left( \frac{Q_b }{Q_a}\right)   \left[f_g(x_a)-x_a f'_g(x_a)\right]\nonumber \\
&\hspace{-1.0cm}+\delta(1-z_a)\log\left( \frac{\Tau}{Q_a}\right)\left[\left( 1-\frac{Q'_a x_a}{Q_a}\right)f_g(x_a) - x_a f'_g(x_a) \right]\Bigg\}.
\end{align}
The total differential cross section is the sum of the two regions
$(a)$ and $(b)$:
\begin{equation}
\label{eq:fullPC2}
\frac{\d\sigma^\text{NLP}}{\d \Tau} =\frac{\d\sigma^{\text{NLP},(a)}}{\d \Tau} +\{a \leftrightarrow b\}.
\end{equation}
This is the full ${\cal O}(1)$ power correction for the $gg$ channel, including both the
LL and NLL terms.  Results for the other channels are given in the Appendix.

Let us first focus on the LL $\log \Tau$ term. Combining both regions we
arrive at the following form valid for arbitrary $Q_a$ and $Q_b$,
subject to the (easily-removeable) restriction that $Q_a Q_b = m_H^2$:
\begin{align}
\label{eq:LLPC}
\frac{\d\sigma^{\text{NLP}}_\text{LL}}{\d \Tau} &= \left(\frac{C_A \alpha_s}{\pi} \right) \int_0^1 \d x_a\int_0^1 \d x_b \frac{(2\pi)}{2 s x_a x_b} \delta(s x_a x_b - m_H^2)  |\mathcal{M}(gg \to H)|^2\nonumber \\
\Bigg\{&\frac{Q_a}{m_H^2}\log\left(\frac{\Tau}{Q_a} \right) f_g(x_b)\left[ \left( 1-\frac{Q'_a x_a}{Q_a}\right)f_g(x_a) - x_a f'_g (x_a)\right]\nonumber \\
+&\frac{Q_b}{m_H^2}\log\left(\frac{\Tau}{Q_b} \right) f_g(x_a)\left[ \left( 1-\frac{Q'_b x_b}{Q_b}\right)f_g(x_b) - x_b f'_g (x_b)\right]\Bigg\}.
\end{align}
We can apply this form to obtain the LL power corrections to the different definitions
of $\Tau$ introduced in Section~\ref{sec:taudef}.  For the fixed
$\Tau$ definition we have 
\begin{equation}
Q_a= Q_b=m_H \qquad \qquad Q'_a = Q'_b =0,
\end{equation}
and therefore the leading-logarithmic power corrections are
\begin{align}
\frac{\d\sigma^{\text{NLP}}_\text{LL}}{\d \Tau} &= \left(\frac{C_A \alpha_s}{\pi} \right) \int_0^1 \d x_a\int_0^1 \d x_b \frac{(2\pi)}{2 s x_a x_b} \delta(s x_a x_b - m_H^2)   |\mathcal{M}(gg \to H)|^2\nonumber \\
\Bigg\{&\frac{1}{m_H}\log\left(\frac{\Tau}{m_H} \right) \left[ 2 f_g(x_a) f_g(x_b) - x_a f'_g (x_a) f_g(x_b)-x_b f_g(x_a)f'_g(x_b)\right]\Bigg\}.
\end{align}
For the hadronic $\Tau$ we have
\begin{equation}
Q_a = x_a \sqrt{s} \qquad Q_b = x_b \sqrt{s} \qquad \qquad  Q'_a = Q'_b = \sqrt{s}.
\end{equation}
The result becomes
\begin{align}
\label{eq:LLhad}
\frac{\d\sigma^{\text{NLP}}_\text{LL}}{\d \Tau} &= \left(\frac{C_A \alpha_s}{\pi} \right) \int_0^1 \d x_a\int_0^1 \d x_b \frac{(2\pi)}{2 s x_a x_b} \delta(s x_a x_b - m_H^2)  |\mathcal{M}(gg \to H)|^2\nonumber \\
\Bigg\{&\frac{\sqrt{s} x_a}{m_H^2}\log\left(\frac{\Tau}{\sqrt{s} x_a} \right) f_g(x_b)\left[  - x_a f'_g (x_a)\right]\nonumber \\
+&\frac{\sqrt{s} x_b}{m_H^2}\log\left(\frac{\Tau}{\sqrt{s} x_b} \right) f_g(x_a)\left[ - x_b f'_g (x_b)\right]\Bigg\}.
\end{align}
These results for the NLO
corrections hold at the level of the inclusive cross section.
\section{Numerical results}
\label{sec:numerics}

We study in this section the numerical impact of the power corrections
computed in the previous section.  Our intent is to compare the
full NLL power
corrections with the LL ones and the hadronic $\Tau$ definition versus the fixed
definition.  We first consider the leading gluon-gluon
partonic channel, before also considering the $qg+gq$ and $q\bar{q}$
results.  The
following parameter choices are used in all results: 
\begin{equation}
\sqrt{s} = 13\, \text{TeV},\;\;\; \mu_R=\mu_F=m_H=125 \,
\text{GeV}.
\end{equation}
All numerical results are obtained using NNPDF 3.0 NNLO parton
distribution functions~\cite{Ball:2014uwa} and MCFM~8.0~\cite{Boughezal:2016wmq}.  We expand the cross section in the
strong coupling constant according to 
\begin{equation}
\sigma_{\text{tot}} = \sigma_{\text{LO}}+\sigma_{\text{NLO}} + \ldots
\end{equation}
and display results for the NLO coefficient $\sigma_{\text{NLO}}$ below.

Finally, we note
that the leading-logarithmic power corrections in Eq.~(\ref{eq:LLPC})
are defined differentially in $\Tau$.  When the below-cut
contribution which includes this power correction is integrated up to
$\Tau^{\text{cut}}$ to obtain the cross section using $N$-jettiness
subtraction, both leading-logarithmic and sub-leading power corrections in
$\Tau^\text{cut}$ are produced:
\begin{equation}
\int_0^{\Tau^\text{cut}} \d \Tau \,  \log\left(\frac{\Tau}{Q} \right) =
\Tau^\text{cut} \left[ \log\left(\frac{\Tau^\text{cut}}{Q}\right)-1\right].
\end{equation}
In our definition of the leading-logarithmic power corrections for the
cross section integrated in $\Tau$ we include both terms arising
from the integral above.

\begin{figure}[h]
  \includegraphics[width=0.7\linewidth]{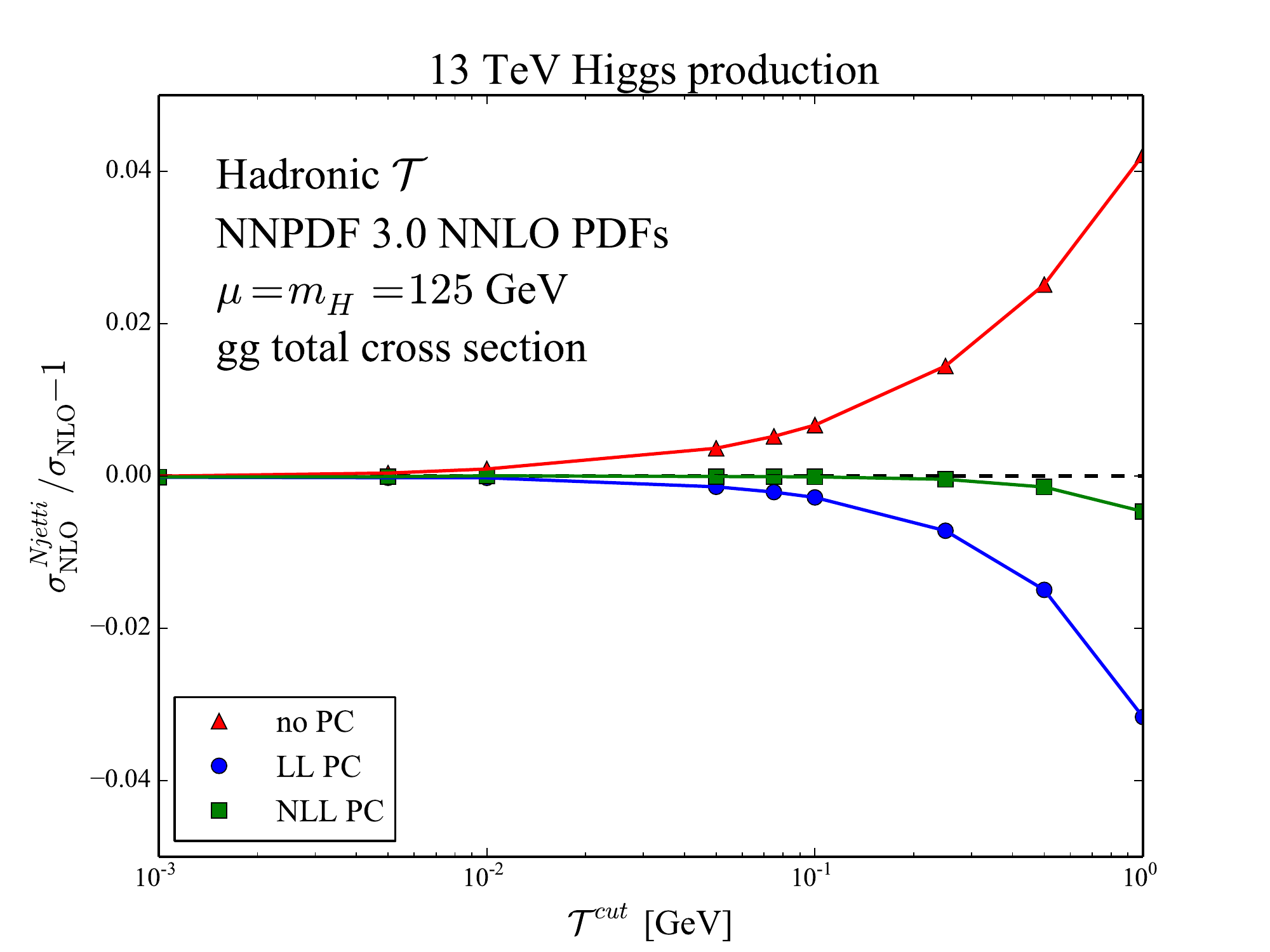}
 \caption{Deviations of the NLO coefficient obtained using $N$-jettiness
subtraction as a function of $\Tau^\text{cut}$  from the dipole-subtraction result for three different
cases: no power corrections included, only the leading-logarithmic
(LL) power corrections included, and the full NLL power
corrections included. }    
 \label{fig:had-PCcomp}
 \end{figure}

We begin by comparing the impact of the power corrections on the
hadronic $\Tau$ distribution.   Shown in Fig.~\ref{fig:had-PCcomp} are the
deviations of the NLO coefficient obtained using $N$-jettiness
subtraction as a function of $\Tau^\text{cut}$  from the dipole-subtraction result for three different
cases: no power corrections included, only the LL power corrections of Eq.~(\ref{eq:LLPC}) included, and the full NLL power
corrections of Eqs.~(\ref{eq:fullPC1}) and~(\ref{eq:fullPC2}) included.  We observe a substantial improvement as first the LL
 and then the NLL power corrections are added to the
leading-power SCET factorization theorem.  The deviation of the NLO
coefficient without power corrections reaches over 4\% at
$\Tau^\text{cut}=1$ GeV.  This is reduced to 3\% with the LL improvements,
and to less than 1\% at NLL.  Requiring a deviation
from the exact result of less than 1\% requires $\Tau^\text{cut} \leq 0.1$
GeV without power corrections.  This is reduced to $\Tau^\text{cut} \leq
0.25$ GeV with LL power corrections included and further to
$\Tau^\text{cut} \leq 1$ GeV with both LL and NLL corrections included.

We now study the impact of power corrections on the fixed $\Tau$
variable.  We also compare the deviations from the dipole subtracton
result between the hadronic and fixed definitions.  We first show in
Fig.~\ref{fig:lephad-noPC} the comparison of the two $\Tau$ definitions without
power corrections for the total cross section.  The fixed definition exhibits less deviation from the
dipole subtraction result for both observables.  Both definitions are
better behaved when the NLL power corrections
are included, as shown in Fig.~\ref{fig:lephad-noPC}.  In particular
the fixed definition becomes nearly identical to the dipole
subtraction result up to $\Tau^\text{cut}=1$ GeV.  This has a significant
effect on the numerical efficiency of the method.  Each of the
above-cut and below-cut contributions depends separately
on $\Tau^\text{cut}$ as $\log^2 (\Tau^\text{cut})$, and these terms only cancel
after the two pieces are added.  We note that with a fixed number of
integrand evaluations for our numerical integration, the estimated
statistical error is a factor of 2 smaller for $\Tau^\text{cut}=1$ GeV than
for $\Tau^\text{cut}=0.05$ GeV, which would have to be used for sub-percent
agreement with the dipole subtraction result for the hadronic
definition without power corrections.

 \begin{figure}
  \includegraphics[width=0.7\linewidth]{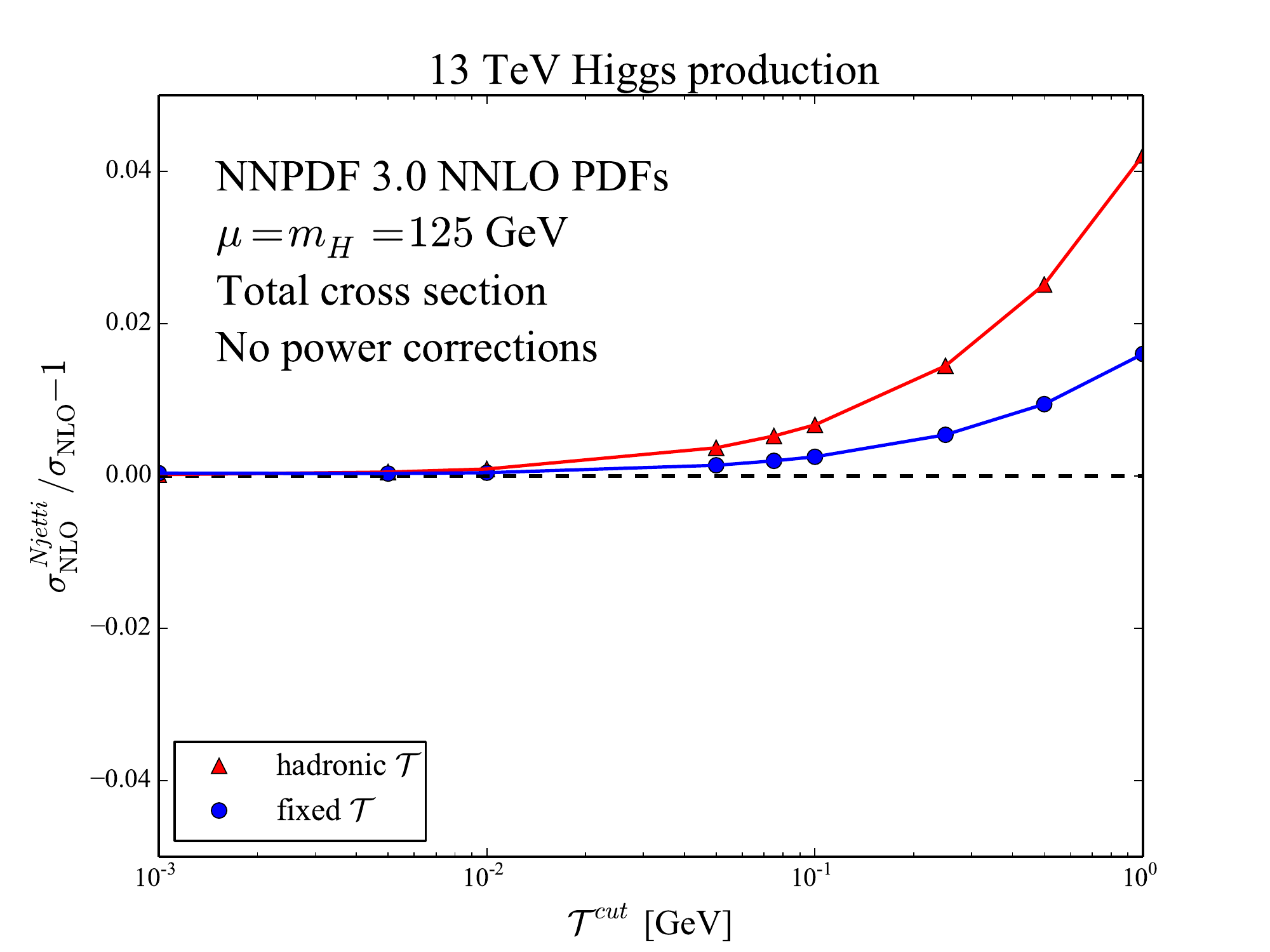}
 \caption{Comparison of the fixed and hadronic $\Tau$ deviations
   from dipole subtraction for the total cross section.  No power corrections
   are included.
   }    
 \label{fig:lephad-noPC}
 \end{figure}

 \begin{figure}
  \includegraphics[width=0.7\linewidth]{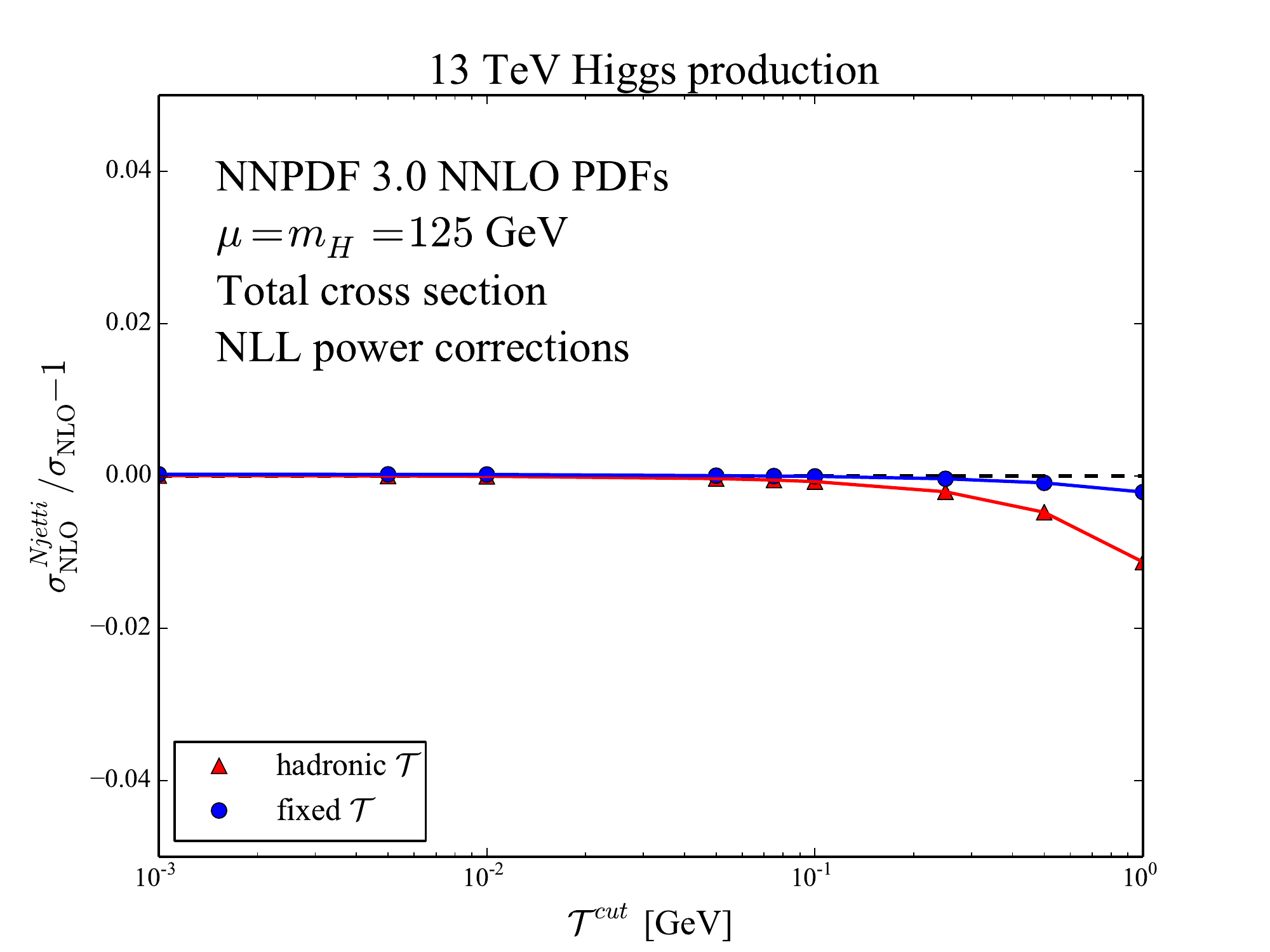}
 \caption{Comparison of the fixed and hadronic $\Tau$ deviations
   from dipole subtraction for the total cross section.  The full ${\cal
     O}(\Tau)$ power corrections are included.
   }    
 \label{fig:lephad-noPC}
 \end{figure}

\begin{figure}
  \includegraphics[width=0.7\linewidth]{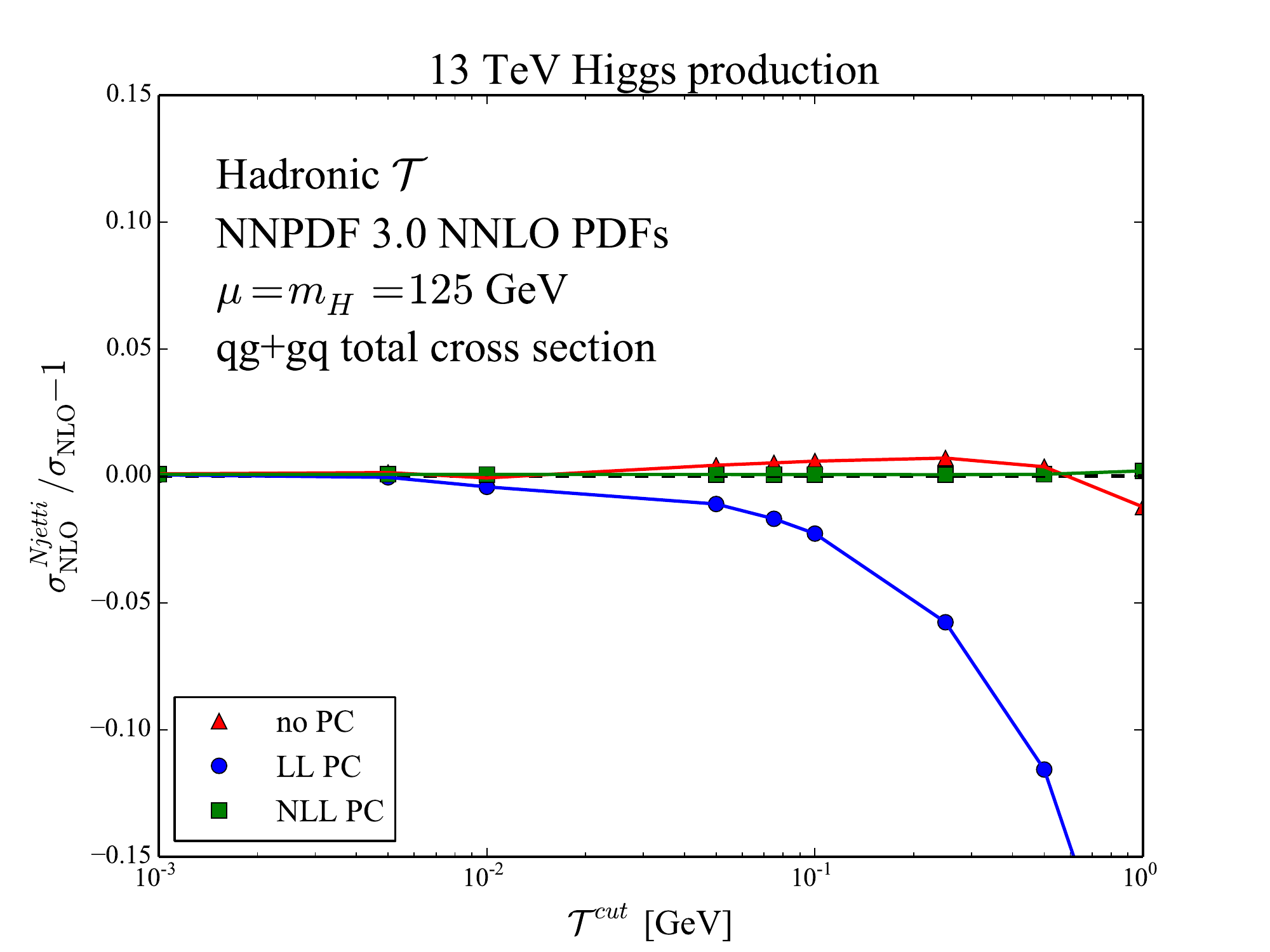}
 \caption{Deviations of the $qg+gq$ NLO coefficient obtained using $N$-jettiness
subtraction as a function of $\Tau^\text{cut}$  from the dipole-subtraction result for three different
cases: no power corrections included, only the leading-logarithmic
(LL) power corrections included, and the full NLL power
corrections included. }    
 \label{fig:qghad-PCcomp}
 \end{figure}

We now consider the $qg+gq$ initial state.  Analytic results for this
channel are given in the Appendix.  We focus on the
hadronic $\Tau$ definition and study the impact of including the full
NLL power corrections.  We note, however, that our derivation is equally valid
for all definitions of $\Tau$.  Results for the inclusive cross section are shown are shown in
Fig.~\ref{fig:qghad-PCcomp}.  It is interesting to note that the LL
power corrections worsen the agreement between $N$-jettiness
subtraction and the dipole subtraction result.  Only upon including
the full NLL power corrections is a better agreement with the dipole
subtraction result obtained.  We note that for the total cross section the agreement at NLL is
excellent, with sub-percent deviations observed all the way up to
$\Tau^\text{cut}=1$ GeV.  The deviation at LL reaches 20\% at this
$\Tauc$ value.  This pattern cannot be seen when studying the full NLO
cross section since the $qg$ channel is much smaller than the dominant
$gg$ scattering process.  We note that the
deviation between the no power-correction result and the dipole
subtraction cross section is non-monotonic as a function of $\Tauc$,
indicating an accidental cancellation between sub-leading powers that
leads to the observed behavior.  The full NLL corrections still
produce a sub-percent deviation from the dipole subtraction result up
to fairly large values $\Tauc \approx 0.3$ GeV.

\begin{figure}
  \includegraphics[width=0.7\linewidth]{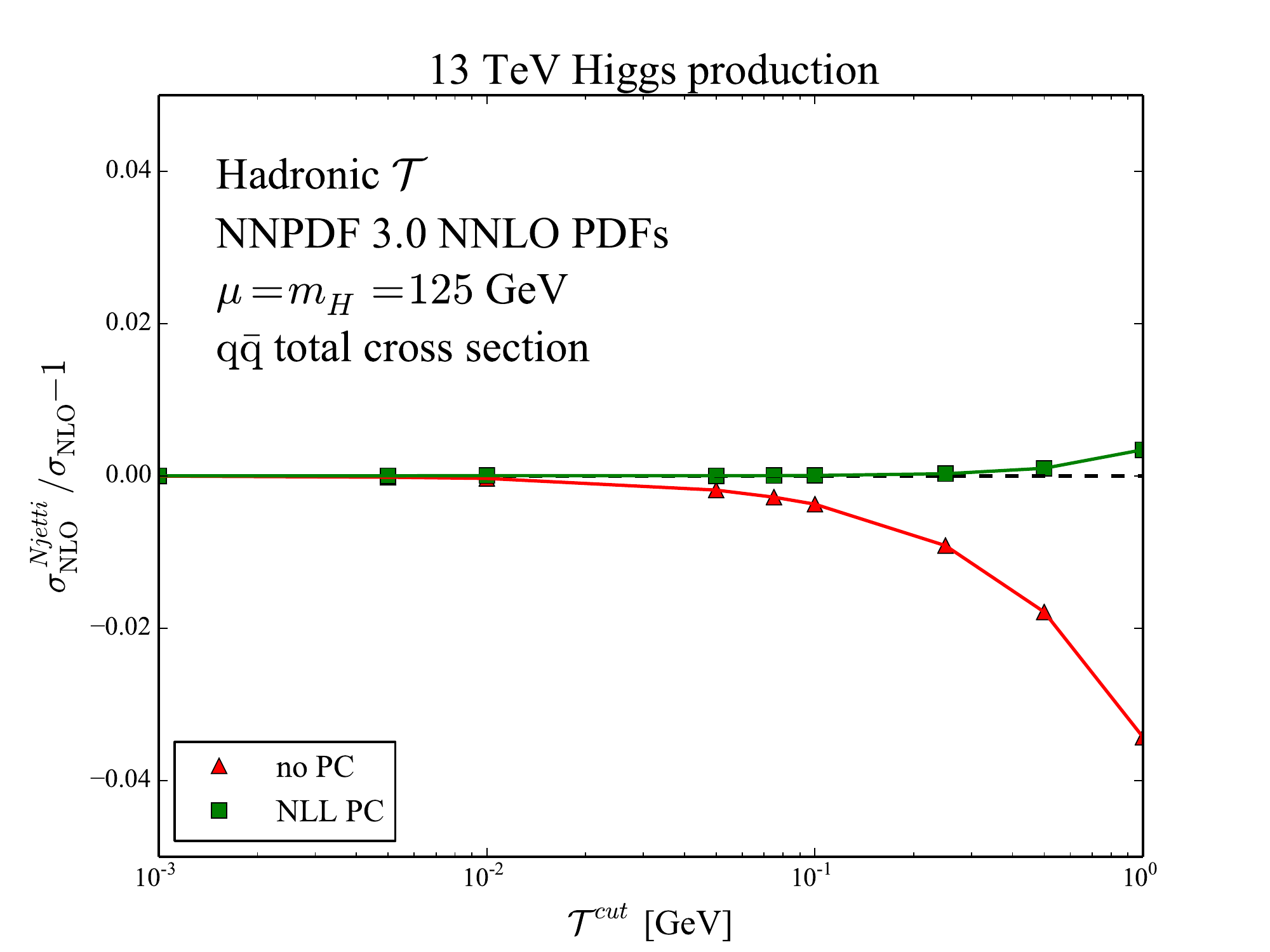}
 \caption{Deviations of the $q\bar{q}$ NLO coefficient obtained using $N$-jettiness
subtraction as a function of $\Tau^\text{cut}$  from the dipole-subtraction result for three different
cases: no power corrections included, only the leading-logarithmic
(LL) power corrections included, and the full NLL power
corrections included. }    
 \label{fig:qqbhad-PCcomp}
 \end{figure}

Finally we consider the $q\bar{q}$ initial state.  Analytic results for this
channel are given in the Appendix.  This
contribution is numerically very subdominant to the other channels.
It does not have either a leading-power or LL sub-leading power
result, and begins first at the NLL level.  Results for the inclusive cross section are shown in
Fig.~\ref{fig:qqbhad-PCcomp}.  The agreement between the dipole and
$N$-jettiness subtraction results after including the NLL power
corrections is excellent, with sub-percent deviations for all studied
values of $\Tauc$. 

\section{Conclusions}
\label{sec:conc}

In this manuscript we have studied in detail the power corrections to
the effective-theory factorization theorem for the 0-jettiness event
shape variable.  In comparison to
previous works we have derived the next-to-leading-logarithmic power
corrections at next-to-leading order, and not just the
leading-logarithmic corrections obtained previously.  We have derived our result directly
in QCD, without use of the effective field theory formalism, and have
discussed the connection between this method and the SCET approach.  We hope that our discussion is of interest to the reader who
desires to better understand the connection between traditional QCD
and the effective-theory approach.  We would like to stress that our result can only be applied to the inclusive cross section, and not to the cross section differential in the Higgs rapidity.  We
have also presented a numerical study of the power corrections for two different $\Tau$ definitions.  Including the NLL
 power corrections improves the performance of the
$N$-jettiness subtraction method applied to color-singlet production,
For the hadronic $\Tau$, including the power corrections reduces the
deviation from the NLO result as computed by dipole
subtraction by nearly a factor of three.  The fixed $\Tau$ shows
nearly no deviation from the dipole subtraction result after the power
corrections are included.  This significantly improves the numerical
efficiency of the $N$-jettiness subtraction method.

It would be interesting in the future to extend the derivation here to
the sub-leading logarithmic level at NNLO as well, which we believe is
possible.  Applications of $N$-jettiness subtraction to jet production
would also
benefit from an understanding of power corrections, which should benefit from the techniques used in our direct
QCD derivation of the 0-jettiness power corrections.  We look forward
to these future investigations.

\section*{Acknowledgments}
 
We thank the authors of Ref.~\cite{Ebert:2018lzn} for discussions that
led to an improved verison of this work.  R.~B. is supported by the DOE contract DE-AC02-06CH11357.  F.~P. is
supported by the DOE grants DE-FG02-91ER40684 and DE-AC02-06CH11357.
A.~I. is supported by the DOE grant DE-FG02-91ER40684 and the NSF
grant NSF-1520916.  This research used resources of the Argonne
Leadership Computing Facility, which is a DOE Office of Science User
Facility supported under Contract DE-AC02-06CH11357. R.~B. and
F.~P. thank the Aspen Center for Physics and the Perimeter Institute
for kind hospitality during the course of this work. This research was supported in part by Perimeter Institute for Theoretical Physics. Research at Perimeter Institute is supported by the Government of Canada through Industry Canada and by the Province of Ontario through the Ministry of Economics Development and Innovation.


\section*{Appendix}
\label{sec:otherchannels}
We compile here the results for the numerically smaller partonic channels,
focusing on the $n$-collinear sector as before.  There are three
channels to consider: $q\bar{q} \to Hg$, $qg \to Hq$, and $gq \to Hq$
(when considering only the $n$-collinear sector the last two channels
are different).  We show the results for each channel in the $n$-collinear sector. The $\overline{n}$-collinear sector is obtained by substituting $a \leftrightarrow b$ (and consequently $gq \leftrightarrow qg$). The full result is the sum of the $n$-collinear and $\overline{n}$-collinear sectors.

\subsection{$q\bar{q} \to Hg$}
There is no leading-power contribution from this channel, as the
matrix element is subleading in $\Tau$.  There is also no
leading-logarithmic power correction. The NLL power correction is
\begin{align}
\label{eq:qqbhg}
\frac{\d\sigma^{\text{NLP},(a)}}{\d \Tau} =& \frac{4}{3}\frac{\alpha_s C_F}{\pi}\int_0^1 \d x_a \int_0^1 \d x_b \frac{(2 \pi) f_{\bar{q}}(x_b)}{2 s  x_a x_b} \delta\left(s x_a x_b - m_H^2\right) |\mathcal{M}(gg \to H)|^2 \nonumber \\
& \times \frac{Q_a}{m_H^2}\int_{x_a  }^{ 1} \frac{\d z_a }{z_a} \frac{(1-z_a)^2}{z_a} f_q\left(\frac{x_a}{z_a} \right).
\end{align}

\subsection{$gq \to Hq$}
There is again no leading-power contribution. The next-to-leading-power has a LL and a NLL term:
\begin{align}
\label{eq:gqhq}
\frac{\d\sigma^{\text{NLP},(a)}}{\d \Tau} =& \frac{\alpha_s C_F}{2\pi}\int_0^1 \d x_a \int_0^1 \d x_b \frac{(2 \pi) f_q(x_b)}{2 s  x_a x_b} \delta\left(s x_a x_b - m_H^2\right) |\mathcal{M}(gg \to H)|^2 \nonumber \\
& \times \frac{Q_a}{m_H^2}\int_{x_a  }^{ 1} \frac{\d z_a }{z_a}
\Bigg\{{-\delta(1-z_a) f_g(x_a)\log\left( \frac{\Tau}{Q_a}\right)+\frac{\mathcal{L}_0 (1-z_a)}{z_a} f_g\left(
  \frac{x_a}{z_a}\right)}\Bigg\}.
\end{align}

\subsection{$qg \to Hq$}
There is a nonzero leading-power contribution that
  comes from the beam function, and that is already in the
  literature. The next-to-leading-power has no leading-logarithmic
  contribution, but it does have a nonzero NLL contribution:
\begin{align}
\label{eq:qghqNLP}
\frac{\d\sigma^{\text{NLP},(a)}}{\d \Tau} =& \frac{\alpha_s C_F}{2\pi}\int_0^1 \d x_a \int_0^1 \d x_b \frac{(2 \pi) f_g(x_b)}{2 s  x_a x_b} \delta\left(s x_a x_b - m_H^2\right) |\mathcal{M}(gg \to H)|^2 \nonumber \\
& \times \int_{x_a  }^{ 1} \frac{\d z_a }{z_a} 
\Bigg\{ \frac{Q_a}{m_H^2}\frac{2-2z_a+z_a^2}{z_a^3} \left[ x_a f_q^{\prime}\left(
    \frac{x_a}{z_a}\right) +x_a z_a \frac{Q_a^{\prime}}{Q_a}  f_q
  \left( \frac{x_a}{z_a}\right)\right]  \nonumber \\
& + \frac{Q_a}{m_H^2}\frac{2-2z_a+z_a^2}{z_a^2}  f_q \left(
  \frac{x_a}{z_a}\right)
-\delta(1-z_a)  \frac{Q_b}{m_H^2} f_q (x_a) \Bigg\}.
\end{align}
%



\end{document}